\documentclass[12pt]{article}

\usepackage{graphicx} % Required for inserting images
\usepackage{amsmath,dsfont,xfrac,latexsym}
\usepackage{amsfonts}

\usepackage[normalem]{ulem}

\usepackage[makeroom]{cancel}

\usepackage{color}
\usepackage{mathtools,slashed}

\definecolor{dgr}{RGB}{0,100,0}

\definecolor{PineGreen}{cmyk}{0.92, 0, 0.59, 0.25}
\definecolor{blue-violet}{rgb}{0.54, 0.17, 0.89}

\definecolor{Gray}{cmyk}{0, 0, 0, 0.50}
\definecolor{MidnightBlue}{cmyk}{0.98, 0.13, 0, 0.43}

\numberwithin{equation}{section}

\begin{document}
%\preprint{...}

\baselineskip=18pt %a la harvmac
\baselineskip 0.6cm
\begin{titlepage}
\vskip 4cm

\begin{center}
\Large{\textbf{Weyl Rescalings: A First Order Approach}}
\par
\end{center}{\LARGE \par}

\begin{center}
\vspace{1cm}
\textbf{L. Andrianopoli}$^{\star,\dagger}$, 
\textbf{J. Zanelli}$^{\sharp,\flat}$ \small
%\\[5mm]
\begin{center}
\today
\end{center}
$^{\star}$Politecnico di Torino, C.so Duca degli Abruzzi 24,
10129 Torino, Italy. \\[2mm]
$^{\dagger}$Istituto Nazionale di Fisica Nucleare (INFN), Section of
Torino, Via P. Giuria 1, 10125 Torino, Italy.  \\[2mm]
$^{\sharp}$Centro de Etudios Cient\'{i}ficos (CECs), Arturo Prat 514,
Valdivia, Chile.  \\[2mm]
$^{\flat}$Facultad de Ingeniería, Universidad
San Sebastián, General Lagos 1163, Valdivia, Chile.
\\[5mm]
\footnotesize \texttt{laura.andrianopoli@polito.it}, 
%\texttt{z@cecs.cl}, 
\texttt{jorge.zanelli@uss.cl}
\par
\end{center}

\vskip 26pt

\begin{abstract}
Weyl transformations, --local rescalings in which the metric transforms as $g_{\mu\nu}(x)\longrightarrow \Omega^2(x)\, g_{\mu \nu}(x)$--, are expected to be a symmetry of nature at extremely high energy, that should be (spontaneously) broken at low energy. In Cartan's  first order formulation of gravity this translates into the condition that the local frames (vielbein) scale as $e^a(x) \longrightarrow \Omega(x) e^a(x)$. In this  construction, the Lorentz connection one-form fields $\omega^a{}_b$ are independent from $e^a$, and since Lorentz transformations and dilatations commute, $\omega$ and $R(\omega)$ must be Weyl invariant, while in the second order metric formalism   the Ricci tensor undergoes a non-homogeneous transformation. This generates a tension between the two formulations, which should be equivalent in the vanishing torsion limit. Here we show how the puzzle can be resolved taking into account the role of torsion. This also sheds some light on the conformal coupling of matter fields to gravity.
\end{abstract}

\end{titlepage}
\newpage
{\baselineskip=12pt \tableofcontents{}}

%%%%%%%%%%%%%%%%%%%%%%%%%%%%%
\section{Introduction}  % 1 %
%%%%%%%%%%%%%%%%%%%%%%%%%%%%%

Nature is not scale-invariant. In our experience, atoms, galaxies, and everything in between, have scales that cannot be changed. Particle masses, ranges of interactions, and fundamental constants like $c$, $e$, $G_{Newton}$ and $\hbar$, that determine the dynamics and the structures that we observe, have fixed values. Nevertheless, it can be expected that at extremely high energies --possibly of the order of the Planck mass--, all phenomena should be scale-invariant: at this scale, all particles are massless, all interactions have infinite range, etc. It is therefore possible to imagine that at extremely high energy, scale invariance could be a good approximate symmetry that is spontaneously broken at lower energies. An argument based on this symmetry has been proposed by 't Hooft as possible resolution to the information paradox \cite{'tHooft2014,tHooft:2015koi}.

In order to understand how scale invariance\footnote{We are not interested here in rigid rescalings that change fields and coupling constants in a manner that is independent of the position in spacetime, as that would be indistinguishable from a change of units of length, time and mass. Assuming that those units have been established, we will consider changing the scales by different factors at different spacetime locations.} could be broken, it is necessary to start by considering a fundamental theory with built-in scale invariance and in particular, a scale invariant theory of gravity, even when coupled to matter sources (typically carrying half-integer spin).

Here, we are not referring to the conformal gravity theory, which is a four-derivative theory for the metric expressed in terms of the Weyl tensor $\mathcal{W}_{ik\ell m}=R_{{ik\ell m}} +\frac 2{n-2}\left(R_{i[m]}g_{\ell]k}-R_{k[m]}g_{\ell]i}
\right) +\frac 2{(n-1)(n-2)}R\,g_{i[\ell]}g_{m]k} $ \cite{Weyl:1918ib,Eisenhart,Carroll,Kaku:1977pa,Mannheim:2011ds,Andrianopoli:2026qsx}, but to a scale-invariant extension of the two-derivative Einstein theory.

Weyl rescalings of the metric\footnote{We use $\Omega(x)=e^{\lambda(x)}>0$ for later convenience.}  are defined as
\begin{equation} \label{Weyl}
g_{\mu \nu}(x)\rightarrow \widetilde{g}_{\mu \nu}(x) = e^{2\lambda(x)}\,g_{\mu \nu}(x)\,,
\end{equation}
together with similar transformations for the matter fields, where $\lambda$ is an arbitrary real, continuous differentiable function. For some physical systems in certain dimensions they form a symmetry group. For example,  the Maxwell action  is Weyl invariant in four dimensions under \eqref{Weyl} if the electromagnetic field is unaltered. This does not happen for the Maxwell action in dimensions other than four.

The Einstein-Hilbert Lagrangian for pure gravity, $L_G(g_{\mu\nu}) = R\sqrt{|g(x)|}$ --where $R$ is the Ricci scalar defined in terms of the metric--, is not invariant under \eqref{Weyl} in any dimension. On the contrary, it transforms inhomogeneously  by a quantity that is not a boundary term \cite{Birrell}
\begin{align} \label{scaling-R0}
R \rightarrow \widetilde{R}= e^{-2\lambda}\left[{R} +(n-1)(2 \Box \lambda + (n-2) g^{\mu\nu} \lambda,_\mu \lambda,_\nu) \right]\,     
\end{align}
(see Appendix \ref{transf app}). However, it was noted long ago that if coupled to a scalar field in a specific way, this Lagrangian becomes Weyl invariant \cite{Chernikov-Tagirov,Freedman-van Proeyen,Carroll}. That is, the action
\begin{equation} \label{Action-s=0}
I[g,\phi]=\int \left(\frac 12\xi\,\phi^2\,R(g(x))+\frac{1}{2}g^{\mu \nu}\partial_\mu \phi \partial_\nu\phi \right)\sqrt{|g|}d^nx\,,
\end{equation}
where $\xi=\frac{n-2}{4(n-1)}$,
 is invariant under \eqref{Weyl}, provided $\phi$ simultaneously transforms as
\begin{equation} \label{s=0}
\phi(x)\rightarrow \widetilde{\phi}(x) = e^{p\lambda(x)}\,\phi(x)\,,
\end{equation}
 for $p=\frac{2-n}{2}$ (see Appendix A, \cite{Eisenhart, Carroll} for details). Then, the field equation for $\phi$,
\begin{align}  \label{eomphi2}  [\Box -\xi R]\,\phi=0\,,
\end{align}
 transforms homogeneously under \eqref{Weyl},\eqref{s=0},
\begin{align}\label{conf_scal}
\left(\Box -\xi \,{R}\right)\phi \longrightarrow & e^{-(\frac{n}{2}+1)\lambda(x)}\left(\Box - \xi\,{R}\right)\phi\,.
\end{align}
This implies that the action 
\begin{equation} \label{Action'-s=0}
I[g,\phi]=\frac 12 \int \sqrt{|g|}d^nx \,\,\phi\,\left[\frac{n-2}{4(n-1)}\,{R} - \Box \right]\phi\,\,,
\end{equation}
--which, up to a surface term, is equal to \eqref{Action-s=0}--, is unchanged under (\ref{Weyl}, \ref{s=0}).

Models with scalar fields conformally coupled to gravity, as in \eqref{eomphi2},  give rise to interesting cosmological consequences. In particular, the spectral features of the cosmic microwave background radiation can be traced to scale invariance of the primordial density fluctuations
\cite{Harrison:1969fb,Zeldovich,Mannheim:2011ds},
 and play  a crucial role in inflationary models \cite{Salopek:1988qh,Bezrukov:2007ep,Tenkanen:2020dge}, also if embedded in supergravity  \cite{Ferrara:2010yw,Ferrara:2010in}. 

The above discussion relies on the so-called \emph{second-order approach} to gravity, where the gravitational dynamics is fully encoded in the metric tensor.
On the other hand, in order to study how Weyl transformations act on  half-integer spin fields, the \emph{first-order approach} is more transparent \cite{Palatini:1919ffw}. In it, the connection is  independent of the metric and the spinor fields couple to the Lorentz connection and not to the Christoffel symbols $\Gamma^\mu_{\nu\rho}= \frac 12 g^{\mu\lambda}\left(\partial_\nu g_{\rho\lambda}+\partial_\rho g_{\nu\lambda} -\partial_\lambda g_{\nu\rho }
\right)$. On top of that, fermionic matter generically produces torsion and therefore assuming vanishing torsion, as usually done in the second order approach, would be contradictory. 

It was recently pointed out that a first-order formulation of   Weyl transformations of a matter-coupled gravity model is inconsistent \cite{Kallosh:2026tjm}. Here we will show how the first and second order approaches can be reconciled if torsion is adequately taken into account.
 
\vskip 5mm

In the first order approach to gravity proposed by Cartan \cite{Cartan}, the Einstein-Hilbert action is a functional of the vielbein one-form $e^a$ (metric structure), and the Lorentz connection one-form $\omega^a{}_b$ (affine structure),
\begin{equation} \label{Ic}
I_C[e, \omega]=\int\epsilon_{a_1 a_2 \cdots a_n}\,R^{a_1a_2}(\omega)\, e^{a_3}\cdots e^{a_n} \,,
\end{equation}
where  $R^{ab}=d\omega^{ab} + \omega^a{}_c\, \wedge \omega^{cb}$ is the Lorentz curvature.\footnote{ In what follows, wedge products between exterior forms are implicitly assumed. }
The  fields $e$ and $\omega$ are assumed to be dynamically independent and therefore independently varied. Under Weyl rescalings \eqref{Weyl}, the vielbein transforms homogeneously while  the Lorenz  connection $\omega$ is Weyl invariant because Lorentz transformations commute with dilatations. Hence, in first order, a Weyl transformation reads
\begin{subequations}\begin{align} \label{W10}
e^a &\longrightarrow \, \tilde e^a\;= e^{\lambda (x)} e^a 
\\
\omega^{ab}  & \longrightarrow \tilde\omega^{ab}  =\, \omega^{ab}\,,\label{omega_tr}
\end{align}
\label{Wfo}
\end{subequations}
and, as a consequence, $R^{ab}(\omega)$ is Weyl invariant as well. Hence, under Weyl rescalings, the first order Einstein-Cartan Lagrangian \eqref{Ic} transforms homogeneously unlike its second order counterpart, $R(g)\sqrt{|g|}$. The inhomogeneous change of $R(g)$ in the second order form is crucial to compensate the inhomogeneous change in the kinetic term for $\phi$ in \eqref{Action-s=0}.

In the absence of torsion, the first- and second order approaches are equivalent in four dimensions. In dimensions greater than four, the natural generalization of Einstein's General Relativity is the Lanczos-Lovelock (LL) theory \cite{Lanczos,Lovelock}, 
$$L\sim \sum_{k=0}^{[\frac n2]} \alpha_k \epsilon \left[R^k\right] \left[e^{n-2k}\right] \,,$$
with $\alpha_k$  arbitrary coefficients, where torsion does not necessarily vanish \cite{Troncoso}. Therefore, in higher dimensions, in the presence of fermionic matter, or for scalar fields non minimally coupled to gravity, the first- and second order approaches can deviate radically. In particular, the conformal coupling $\phi^2\, R(g)$ in second order and $\phi^2 \epsilon_{abc_3 ...c_n} R^{ab} e^{c_3}...e^{c_n}$ in first order transform differently and give rise to very different field equations. Here we examine this issue and show that, taking into account the change in torsion induced by the Weyl transformations, the two approaches are consistent.

Black hole solutions containing conformally coupled scalar fields at first order have been recently discussed in \cite{Aviles:2024muk,Aviles:2026xzt}.

The paper is organized as follows. Section 2 reviews the splitting between torsion-free and torsionful parts of the Lorentz connection. This is  used to split a naturally Weyl invariant first order action  into pieces that can be confronted with the second order expressions. Section 3 contains the extension to fermionic matter and to Maxwell/Yang-Mills fields. Section 4 contains a discussion and summary.

%%%%%%%%%%%%%%%%%%%%%%%%%%%%%%%%%%%%%%%%%%
\section{First order formulation}  % 2 %
%%%%%%%%%%%%%%%%%%%%%%%%%%%%%%%%%%%%%%%%%%

Weyl transformations are associated with a $SO(1,1)$ group and, in a Lorentz covariant theory, this factor commutes with the Lorentz group $SO(n-1,1)$.\footnote{Lorentz and Weyl groups get unified in the conformal group $SO(n,2)$, which admits as a maximal semisimple subgroup $ SO(1,1)\times SO(n-1,1)$. The  conformal group $SO(n,2)$ emerges indeed as a hidden symmetry of the Weyl theory of conformal gravity \cite{Kaku:1977pa}. A  first-order formulation of conformal gravity was recently constructed in \cite{Andrianopoli:2026qsx}.} A Weyl transformation \eqref{Weyl} clearly affects the vielbein 1-form, since $e^a_\mu$ are ``the square root" of the metric. The Lorentz connection, on the other hand, is an $SO(n-1,1)$-gauge field that relates the local orthonormal bases in the cotangent bundle. Since the local Minkowski spaces tangent to the spacetime manifold are scale invariant, Weyl and  Lorentz transformations must commute and therefore the Lorentz connection should not be charged with respect to the Weyl group $SO(1,1)$. Consequently, \eqref{Weyl} translates in  first order language into the form given by \eqref{Wfo}. For the inverse vielbein $E^\mu_a$, Weyl transformations act as
\begin{equation}
\label{W10_inv}
E^\mu_a \longrightarrow \tilde{E}^\mu_a = e^{-\lambda(x)}\,E_a^\mu\,.
\end{equation}
As a consequence of \eqref{omega_tr}, the Lorentz covariant derivative, $D=d+\omega$ and the corresponding curvature, $R^a{}_b=d\omega^a{}_b + \omega^a{}_c   \omega^c{}_b$, are both Weyl invariant. This contrasts with the second-order formulation, in which the Weyl rescaling of the metric \eqref{Weyl} fixes the transformation of the Christoffel symbols and, consequently, the Riemann curvature transforms as (see Appendix \ref{transf app})
\begin{align} \label{Riem-tr}
R^\mu{}_{\nu \alpha \beta} \to \widetilde{R}^\mu{}_{\nu \alpha \beta} =&  R^\mu{}_{\nu \alpha \beta}\, + \delta^\mu_\beta \lambda_{\nu \alpha} - \delta^\mu_\alpha \lambda_{\nu \beta} +g^{\mu \sigma}(g_{\nu\alpha}\lambda_{\sigma \beta}- g_{\nu\beta}\lambda_{\sigma \alpha}) \nonumber\\ & + (\delta^\mu_\beta g_{\nu\alpha} - \delta^\mu_\alpha g_{\nu\beta})  
g^{\sigma\tau}\lambda_{,\sigma}\lambda_{,\tau} \,.
\end{align}
Nevertheless, the first and second order formulations should be equivalent if and only if the torsion vanishes. 

\vskip 5mm

In general, the Lorentz connection $\omega^{ab}$ and the vielbein $e^a$ are related by the torsion two-form, 
\begin{equation} \label{T}
T^a=De^a = de^a+\omega ^a{}_b e^b\,,
\end{equation}
which can be seen as an algebraic equation for $\omega$.  The solution $\omega=\omega(e, \partial e, T)$ has the form
\begin{equation} \label{omegasplit}
\omega^a{}_b=\mathring{\omega}^a{}_b+\kappa^a{}_b\,,
\end{equation}
where $\mathring{\omega}^a{}_b$ is the torsion-free part, defined by the condition $\mathring{D}e^a\equiv de^a+\mathring{\omega}^a{}_b e^b = 0$,
\begin{equation} \label{tor_free}
\mathring{\omega}^a{}_{b\mu}= e^a_\lambda\left(\partial_\mu E^\lambda_b + \Gamma^\lambda_{\mu \nu} E^\nu_b \right)= e^a_\lambda \nabla_\mu E^\lambda_b\,,
\end{equation}
and  $\nabla_\mu$ is the covariant derivative on the spacetime manifold for the Christoffel symbol. The contorsion $\kappa^a{}_b$ in \eqref{T} is a one-form such that
\begin{equation} \label{kappa}
T^a=\kappa^a{}_b e^b\,.
\end{equation}

Under the Weyl transformation \eqref{Wfo}, the torsion two-form changes as
\begin{align}
T^a\rightarrow \tilde T^a &= e^{\lambda(x)}\,(T^a+ d\lambda\, e^a)\nonumber\\
&=  e^{\lambda(x)}\,(\kappa^a{}_b\, e^b+ d\lambda\, e^a)\,.\label{k0}
\end{align}
Comparing this with $\tilde\kappa^a{}_b\, \tilde e^b$, leads to
\begin{align} \label{tildek-k}
\tilde \kappa^a{}_b e^b = \kappa^a{}_b e^b + d\lambda\, e^a\,. 
\end{align}
Since $\kappa^{ab}=-\kappa^{ba}$, eq. \eqref{tildek-k} can be solved for the contorsion, resulting in the following expression for the transformed contorsion
\begin{align} \label{tildek-mu}
\tilde \kappa^a{}_{b\mu} &= \kappa^a{}_{b \mu} + (e_{b\mu}E^{\rho a} - e^a_\mu \,E^\rho_b) \partial_\rho\lambda\,, \end{align}
or
\begin{align} \label{tildek-ab}
\tilde \kappa^a{}_b = \kappa^a{}_b+ \left(\eta^{ad} \eta_{bc} - \delta^a_c \delta^d_b \right) e^c \,\partial_d\lambda\,, 
\end{align}
where $\partial_a \equiv E^\mu_a \partial_\mu$. 
From the last expression we see, in particular, that starting from a torsion-free geometry, a Weyl transformation produces a contorsion tensor 1-form given by
\begin{align} \nonumber
\tilde \kappa^a{}_b(\lambda) &= \tilde  e_b\tilde\partial^a \lambda - \tilde e^a \tilde \partial_b \lambda \\ \nonumber
&= \tilde e_b \tilde \partial^a \lambda - \tilde e^a \tilde \partial_b \lambda   \\ \label{k} &= \left( \eta^{ac}\eta_{bd} -\delta^a_d \delta^c_b  \right)\tilde e^d \tilde \partial_c \lambda\,, 
\end{align}
where $\partial^a= \eta^{ab}\partial_b$ (note that since $E^\mu_a$ and $e^a_\mu$ scale inversely, the tildes can be dropped from the product $\tilde{e}^d \tilde{\partial}_c$).

In the second order formulation, on the other hand, a Weyl rescaling of the metric in a torsion-free geometry does not generate torsion, because the transformed affine connection is forced to remain symmetric in its lower indices. Therefore, Weyl rescalings are very distinct operations in the first- and second-order formalisms.

If the starting geometry has no torsion ($\omega=\mathring \omega$), the Weyl transformation \eqref{W10} generates a nonvanishing torsion 2-form in the rescaled geometry,
\begin{equation}
T^a=D e^a=0\longrightarrow \tilde D \tilde e^a = D (e^{\lambda(x)} e^a) = d\lambda \tilde e^a\,.
\end{equation} 
This means that the ``new" Lorentz connection, $\tilde{\omega}^{ab}$ --which is identical to the original one--, is now torsionful with respect to the new vielbein $\tilde{e}^a$. 
Then, under a Weyl transformation, the purely metric Riemann-curvature 2-form $\mathring R\equiv R(\mathring\omega)$ transforms \emph{inhomogeneously} as
\begin{align}  \mathring{R}^{ab}\longrightarrow  \tilde R^{ab} =\mathring{R}^{ab}=\mathring{\tilde R}^{ab} + \mathring{\tilde D}\kappa^{ab}(\lambda)+ \kappa^a{}_c (\lambda) \kappa^{cb}(\lambda)\,.\label{r0trasf}
\end{align}
Using the definition $R^{ab}=\frac 12 R^{ab}{}_{\mu\nu}dx^\mu\wedge dx^\nu$, and eq. \eqref{tildek-ab}, then \eqref{scaling-R} follows directly, since $\mathring R= R(g)$. In other words, while the Lorentz curvature is invariant under a Weyl rescaling, its torsion-free part transforms inhomogeneously, in correspondence with the second order form.

%%%%%%%%%%%%%%%%%%%%%%%%%%%%%%%%%%%%%%%%%%%%%%%%%%%%%%%%
\subsection{First order Weyl invariant gravity} % 2.1%
%%%%%%%%%%%%%%%%%%%%%%%%%%%%%%%%%%%%%%%%%%%%%%%%%%%%%%%

 In first order form, the $n$-dimensional Einstein-Hilbert action  with cosmological constant is  
\begin{align} \label{EH}
I_{EH} [e,\omega]= \frac{-1}{(n-2)!}\int \epsilon_{a_1 \cdots a_n} \left( R^{a_1a_2}+ \Lambda e^{a_1} e^{a_2} \right)\,e^{a_3}\cdots  e^{a_n}
\end{align}
which is equivalent to its second order counterpart, given in terms of the Einstein-Hilbert Lagrangian, $\sqrt{|g|}(R(g) + n(n-1)\Lambda)$ if $\omega$ is torsion-free.

This action is not Weyl invariant, but it can be modified to a Weyl invariant one  (without higher derivative interactions as in conformal gravity) if a non minimally coupled scalar field in included. The action
\begin{align}\label{confscalar1}
I_0[e,\omega, \phi]=\frac{-2}{(n-2)!}\int \,\left(\frac{1}{2} \phi^2 R^{ab}(\omega)+ \frac{1}{q}\phi^q \,e^a\, e^b \right)\,e^{c_3} \cdots e^{c_n}\epsilon_{abc_3\cdots c_n}
\end{align}
is trivially Weyl invariant if  $q=2n/(n-2)$ and torsionful, provided $\phi$ scales as in \eqref{s=0}. Varying with respect to the connection gives an equation for torsion
\begin{equation} \label{dphi}
\epsilon_{abc_3\cdots c_n} \left[(d\phi^2)\,e^{c_3} \cdots e^{c_n}+\phi^2 D(e^{c_3} \cdots e^{c_n}) \right]=0\,,
\end{equation}
whose solution is
\begin{equation}
T^a = \frac 12 T^a{}_{\mu\nu}dx^\mu dx^\nu=\frac 2{n-2}e^a d\ln\phi\,,
\end{equation}
which is an algebraic equation for $\omega$ and can therefore be legitimately substituted in $I_0$ to produce an equivalent dynamical theory.\footnote{This is a consequence of the implicit function theorem extended to functional analysis \cite{Henneaux}.}

From the definition of torsion, the Lorentz connection can be written as 
\begin{equation}\label{omega(e,fi)}
\omega^{ab}\equiv \mathring{\omega}^{ab}(e)+\kappa^{ab}(\phi,e)\,,
\end{equation}
where $\mathring D e^a=d e^a+ \mathring{\omega}^a{}_b\, e^b \equiv 0$ defines the torsion-free part of the Lorentz connection and the contorsion,
\begin{align}\label{T-free.omega}
\mathring{\omega}^a{}_{b\mu}(e) =& \,e^a_\nu \,\left( \partial_\mu E^\nu_b + \mathring{\Gamma}^\nu_{\mu \lambda} E^\lambda_b \right)\\
\label{K(fi)}
\kappa^a{}_{b\mu}(\phi,e) =& \frac 2{n-2} (\partial_d\ln\phi)\,\left(\delta^a_c \delta^d_b -\eta_{bc} \eta^{ad}\right)\, e^c_\mu \,.
\end{align}
The Lorentz curvature then splits into the Riemann curvature $\mathring{R}$ and terms involving the contorsion,
\begin{align}
R^{ab}(\omega)= \mathring{R}^{ab}(\mathring\omega(g))+ \mathring D\kappa^{ab}(\phi,e) + \kappa^a{}_c(\phi,e) \,\kappa^{cb}(\phi,e)\,.
\end{align}
Substituting this expression in the action yields
\begin{align} \nonumber
I_0[\mathring\omega(e), \phi] =\frac{-1}{2(n-2)!}\int 
\,\phi^2&\Bigl( \mathring R^{ab}(g) +\kappa^a{}_c \, \kappa^{cb} +\mathring D \kappa^{ab} + \frac{1}{q}\phi^{q-2} \,e^a\, e^b \Bigr)\times \\ \label{i0g}
& \times e^{c_3} \cdots e^{c_n}\epsilon_{abc_3\cdots c_n}\,.
\end{align}
Up to a total derivative, the third term in \eqref{i0g} can be rewritten as
$$\mathring D \kappa^{ab}e^{c_3} \cdots e^{c_n}\epsilon_{abc_3\cdots c_n}=-2\phi \, d\phi \, \kappa^{ab}\, e^{c_3} \cdots e^{c_n}\epsilon_{abc_3\cdots c_n}\,,$$ 
so that, using \eqref{K(fi)}, the two contorsion terms in \eqref{i0g} sum up and give (recalling that $\xi= \frac{n-2}{4(n-1)}$)
\begin{align}
I_0[e,\mathring\omega, \phi] 
=\int \left(\frac 12\xi\phi^2 \mathring R+ \frac 12\partial_\mu\phi\partial^\mu\phi + \frac{1}{q}\phi^{q-2} \,e^a\, e^b\right) |e| d^nx \,,  \label{ghost}
\end{align}
which coincides with \eqref{Action-s=0} with the addition of the scalar self-coupling term. 

This analysis shows that the first order Weyl invariant action \eqref{EH} and the second order Weyl invariant action \eqref{Action'-s=0} are indeed equivalent. The fact that in \eqref{EH} torsion is not zero, merely reflects the fact that $\phi$ is not constant.

%%%%%%%%%%%%%%%%%%%%%%%%%%%%%%%%%%

%%%%%%%%%%%%%%%%%%%%%%%%%%%%%%%%%%%%%%%%%%%%%%
\section{Spinor and vector fields}
%%%%%%%%%%%%%%%%%%%%%%%%%%%%%%%%%%%%%%%%%%%%%%
Next we examine Weyl rescalings for spin-1/2, spin-1 and spin-3/2 fields coupled to Einstein-Cartan gravity.\\

\textbf{Spinors}\\
The first order formulation is the natural setup to describe fermions in a curved background. In this setting, the $n$-form Lagrangian for a spin-1/2 \textbf{Dirac field} $\chi$ minimally coupled to gravity is given by
\begin{align} \label{Dirac-L0}
\mathcal{L}_D= &  -\frac 1{2(n-1)!}\left(\bar\chi \gamma^a \nabla \chi - \nabla\bar\chi \gamma^a \chi\right) e^{b_1} \cdots e^{b_{n-1}}\epsilon_{a b_1\cdots b_{n-1}}\end{align}
where $\nabla=d+\frac{1}{4}\,\omega^{ab}\,\gamma_{ab}$ is the Lorentz-covariant exterior derivative for a generic (possibly torsionful) background $\omega$, in the spin-1/2 representation. This Lagrangian is invariant under a Weyl rescaling \eqref{Wfo}, provided the spinor transforms as 
\begin{align}\chi \to \tilde \chi= e^{p\lambda} \chi\,, \quad \bar{\chi} \to \tilde{\bar{\chi}} = e^{p\lambda} \bar{\chi}\,, \text{ with: } p= \frac{1-n}2\,.\label{spin_tr0}\end{align}

We remark that Weyl invariance of the Dirac Lagrangian \eqref{Dirac-L0} is  automatic because of three features of the first order approach: i) the Lorentz connection is Weyl-invariant; ii) the conformal weight of the fermionic bilinears is $n-1$ and there are exactly $n-1$ vielbein in $\mathcal{L}_D$; and iii) the derivatives of $\lambda$ cancel out in the Lagrangian. So, unlike a pure scalar field, a pure Dirac field does not require the presence of gravity to achieve Weyl invariance.

In the second order formulation, on the other hand, Weyl invariance depends on how the Lagrangian is defined. One possibility would be to define
\begin{align}
 \mathring{\mathcal{L}}_D= \frac 12 \left(\bar\chi \gamma^a\mathring{\nabla}_\mu \chi - (\mathring{\nabla}_\mu\bar\chi) \gamma^a \chi\right)E^\mu_a \sqrt{-g}d^nx\label{Dirac-L2}
\end{align} 
where $\mathring{\nabla}$ is the  covariant derivative in the Lorentz representation for the torsion-free connection $\mathring{\omega}$. This Lagrangian can describe a spinor field defined on a prescribed non-dynamical geometry whose metric and connection are given at each spacetime point. This would be compatible with the torsion-free condition for (3.3), because it only defines the propagation of $\chi$ on that background.

Another possibility would be to start from the first order Lagrangian \eqref{Dirac-L0} replacing the Lorentz connection by its split form $\omega=\mathring{\omega}+\kappa$, where $\mathring{\omega}^a{}_{b\mu}= e^a_\nu (\partial_\mu E^\nu_b+ \Gamma^\nu_{\mu \lambda}\, E^\lambda{}_b)$. In that case, Weyl invariance of \eqref{Dirac-L2} is nontrivial since, if one uses \eqref{T-free.omega}, a subtle cancellation takes place between the terms $\nabla \lambda$ in $\nabla \tilde \chi$ with those in $\tilde \nabla_\mu \tilde E^\mu_a$, in which the metric condition $\partial_\mu E^\nu{}_a-\omega_a{}^b\,E^\mu{}_b + \Gamma^\mu_{\nu \lambda}\, E^\lambda{}_a=0$ should be used,  (where $\Gamma$ is any metric-compatible affine connection, torsionful or not).

 A dynamical Weyl-invariant geometry, on the other hand, means that the action should also include a gravitational term in the Jordan frame, as in \eqref{Action-s=0}. This is equivalent to the first order expression $\epsilon_{ab c_3\cdots c_n}\,\phi^2 \,R^{ab}\,e^{c-3} \cdots e^{c_n}$, where the non minimally coupled scalar $\phi$ gives an additional contribution to the torsion, as we have seen. In sum, for a Dirac action coupled to dynamical Weyl-invariant gravity action, torsion would involve derivatives of $\phi$ and bilinears of the form $\bar{\chi} \gamma \chi$. When that expression for torsion is substituted in the second order Lagrangian, it would introduce couplings between the fermion and the scalar and quartic fermionic terms as well. At any rate, this would be inconsistent with the assumption that the connection in the Lagrangian is torsion-free and $\omega=\mathring{\omega}$. 

\vskip 5mm

The first-order \textbf{Rarita-Schwinger} Lagrangian in $n$ dimensions,
\begin{equation}
\mathcal L_{RS}=\frac{-1}{12 (n-3)!} \left(\bar\psi \gamma^{a_1 a_2 a_3}\nabla\psi -\nabla\bar\psi \gamma^{a_1 a_2 a_3}\psi\right) e^{a_4}\cdots e^{a_n}\epsilon_{a_1 \cdots a_n}
\end{equation}
is Weyl invariant provided the gravitino 1-form field $\psi$ rescales as
\begin{align}
\psi \longrightarrow e^{p\lambda} \psi\,, \text{ with: } p=\frac{3-n}2\,.
\end{align}
The invariance of the first-order action is automatic also in this case, along the same lines discussed for the spin-1/2 fermions.

\vskip 5mm

\textbf{Gauge vectors}\\
First order Lagrangians for Maxwell and Yang-Mills fields can   be written at first order in a geometric way   by introducing tensorial auxiliary fields (see \cite{Castellani:1991eu}, Chapter II.9). The one-form gauge connection fields $\mathbb{A}=\mathbb{A}_\mu dx^\mu$ are coordinate-frame independent and Lorentz scalars. Therefore they do not couple neither to the Christoffel symbol nor to the Lorentz connection --they are uncharged with respect to diffeomorphisms and spacetime symmetries. Consequently, they do not need the gravitational Lagrangian in order to compensate for their scale transformation to define a Weyl invariant action.
In the first-order description, this is confirmed by writing the Lagrangian for a spinor minimally coupled to the gauge field,
$$-\frac 1{2(n-1)!} \,\overline \chi \gamma^a \chi \,\mathbb{A} e^{c_1}   \cdots e^{c_{n-1}}\epsilon_{ac_1\cdots c_{n-1}}+ \mathcal{L}_D\,,$$
where 
$\mathcal{L}_D$ is the 1st-order Dirac Lagrangian given in \eqref{Dirac-L0}. This  is Weyl invariant for $\chi \to e^{u\lambda}\chi$, with $u= \frac{1-n}2$, if the 1-form gauge connection $\mathbb{A}$ (and therefore also its field-strength $F=\frac 12 F_{\mu\nu} dx^\mu dx^\nu$) is Weyl invariant.

Additional constraints on the  Weyl transformation of gauge fields arise when writing the standard metric-dependent gauge-field kinetic terms, 
\begin{align}
-\frac 14 F_{\mu\nu}F_{\rho\sigma} g^{\mu\rho}g^{\nu\sigma}\sqrt{|g|}\,.
\label{kinA}
\end{align} At first order, this is obtained as
\begin{align}
   \label{Lgauge0} \mathcal{L}=& \frac 1{2(n-2)!} F \mathcal{F}^{ab} e^{c_1}\cdots   e^{c_{n-2}}\epsilon_{abc_1 \cdots c_{n-2}} +\nonumber\\
   &-\frac 1{4\,n!}\mathcal{F}^{ab}\mathcal{F}_{ab}e^{c_1}\cdots  e^{c_{n}}\epsilon_{c_1 \cdots c_{n}  },
\end{align}
 where $\mathcal{F}_{ab}$ is an auxiliary tensorial 0-form, useful to write the kinetic term at first-order, without introducing the Hodge-operator  (which is explicitly dependent on the metric). The field equation for the auxiliary field gives
\begin{align}  \mathcal{F}_{ab} = F_{ab}= F_{\mu\nu} E^\mu_aE^\nu_b\,.\label{Ff0}
\end{align}
 Then,  the first term is invariant if the auxiliary 0-form transforms as $\mathcal{F}_{ab}\to e^{(2-n)\lambda}\mathcal{F}_{ab}$.  On the other hand, the second term is instead  invariant if $\mathcal{F}_{ab}\to e^{-n\lambda/2}\mathcal{F}_{ab}$. These two transformation laws are compatible only for $2-n=-n/2$, that  is for $n=4$, in which case they give $\mathcal{F}_{ab}\to e^{-2\lambda}\mathcal{F}_{ab}$, which is compatible with \eqref{Ff0}, taking into account the  Weyl transformation of the inverse vielbein, eq.\eqref{W10_inv}. 

Note that the above restriction does not apply to the Chern-Simons theories –gravity included–, whose only dynamical
fields are scale-invariant
gauge connections and do not have a metric-dependent kinetic term of the form \eqref{kinA}.

%%%%%%%%%%%%%%%%%%%%%%%%%%%%%%%%%%%%%%
\section{Discussion and summary} % 4 %
%%%%%%%%%%%%%%%%%%%%%%%%%%%%%%%%%%%%%%

The first order formulation of Weyl-invariant Lagrangians for fields of spin $s=0,\, 1/2, 1, 3/2$ and 2 has been presented. The approach is straightforward to implement, and always consistent in the handling of fermions. The scalar field ($s=0$) and gravity ($s=2$) need each other in a delicate non-minimal coupling to produce a Weyl invariant system. 
The only exception are the Chern-Simons gravity theories, which are scale invariant by construction.

\vskip 5mm
The gauge fields --Lorentz and internal-symmetry  connection one-forms-- are Weyl invariant, while the vielbein,  scalars and fermionic fields transform in a definite way in each dimension. Since the Chern-Simons forms are defined by the connection and independent of the metric structure, they are automatically Weyl invariant always. The Maxwell/Yang-Mills fields with metric-dependent kinetic terms define Weyl invariant systems if and only if the spacetime dimensions is four.

Dirac and Rarita-Schwinger   actions, on the other hand, are Weyl invariant in any dimension, with conformal weight $p=s-n/2$. For these fields, the second order formulation can always be consistently defined on a non-dynamical background spacetime. If coupled to a dynamical gravitational system, a conformally coupled scalar field would be required for Weyl invariance, which would bring in additional Yukawa couplings $\sim \bar{\psi}\phi \psi$, and quartic Nambu--Jona-Lasinio terms $\sim (\bar{\psi}\gamma\psi)^2$.

\vskip 5mm
The first order action \eqref{confscalar1} is naturally obtained in the unconventional supersymmetric (USUSY) extension \cite{Alvarez:2021zhh,ususy} of the AdS algebra in four dimensions. This generalizes the identification  \cite{Townsend:1977xw} between the gauge field associated with AdS boosts  and the  vielbein, $f^a= \phi e^a$. In this sense, the Weyl-invariant action \eqref{Action-s=0} can be seen as the gravitational sector of the four-dimensional USUSY theory. The Weyl invariance of \eqref{confscalar1} is inherited from the USUSY theory, essentially a Yang-Mills system for an $osp(2|4)$ connection, which is Weyl invariant for any gauge (super) algebra in four dimensions.
\vskip 5mm
The Weyl-invariant scalar-tensor Lagrangian --c.f., \eqref{Action-s=0}, \eqref{ghost}-- presents a puzzle: either the scalar field or the gravity sector have the wrong sign for the kinetic term. There seem to be two viewpoints about this issue: i) It describes an ordinary scalar field conformally coupled to a non-dynamical curved background, where the Ricci scalar acts as a locally defined addition to the mass; ii) It describes a gravitational action conformally coupled to a ghost scalar field. The first picture is that of Faraoni and Sonego \cite{Faraoni-Sonego}, Breitenlohner and Freedman \cite{Breitenlohner-Freedman}, for example. In particular, this framework is the correct one to analyse the propagation of a scalar field on a constant curvature background.
The second viewpoint is that of Eisenhart \cite{Eisenhart}, Freedman and van Proeyen \cite{Freedman-van Proeyen} and Carroll \cite{Carroll}. 
 The fact that in this case the scalar field is a ghost  at first discouraged further investigations towards a realistic cosmological model, but was   then exploited to build inflationary models \cite{Salopek:1988qh},\cite{Futamase:1987ua}.
The two viewpoints were  integrated together in 
\cite{Ferrara:2010in},  to build a phenomenological model in the context of the next-to-minimal supersymmetric standard model, able to implement an Higgs-type inflation. In the Weyl invariant model considered,  two scalar fields, both non minimally coupled to gravity, are required to trigger inflation and studied in the Jordan frame. One of them is a ghost, while the other has a canonical kinetic term.

\vskip 5mm
The first order action \eqref{confscalar1} can be seen as a way to promote Newton's and cosmological constants to some features of the dynamical scalar field, $\phi^2\sim G^{-1}$ and $\phi^k \sim \Lambda$. This is an attractive idea in its own right that might be independently pursued. 

Here, we did not consider conformal gravity, which is a higher derivative theory, quadratic in the Riemann tensor. For that theory, a hidden invariance under  the full  conformal group shows up  besides the Weyl symmetry \cite{Kaku:1977pa}, and it becomes manifest in the  first-order formulation of the theory, as shown in  \cite{Andrianopoli:2026qsx} for the $n=4$ case. In that case, the Weyl symmetry is a gauge symmetry which can be expressed as a trace part in a generalized spin connection \cite{Weyl:1918ib}, an exact part of which can be identified as non-metricity. 

The first order formulation presented here is a way to circumvent the potential obstructions to couple conformal gravity at first order to matter, pointed out  in  \cite{Kallosh:2026tjm}. 

\appendix
%%%%%%%%%%%%%%%%%%%%%%%%%%%%%
\section{Weyl transformations in the second order approach}\label{transf app}
%%%%%%%%%%%%%%%%%%%%%%%%%%%%%

Under a Weyl transformation \eqref{Weyl}, the Christoffel symbols, $$\Gamma^\alpha_{\beta\gamma}=\frac{1}{2}g^{\alpha\mu}(g_{\mu\beta},_\gamma + g_{\mu\gamma},_\beta - g_{\alpha\beta},_\mu)$$
and the (torsion-free) Riemann curvature, $$R^\mu{}_{\nu \alpha \beta} = \partial_\alpha \Gamma^\mu_{\nu \beta}  -\partial_\beta \Gamma^\mu_{\nu \alpha} + \Gamma^\mu_{\rho \alpha}\Gamma^\rho_{\nu \beta} - \Gamma^\mu_{\rho \beta}\Gamma^\rho_{\nu \alpha}\,,$$ transform as 
\begin{align}\label{Gamma}
\Gamma^\alpha_{\beta \gamma} \;\longrightarrow \widetilde{\Gamma}^\alpha_{\beta \gamma} =
\Gamma^\alpha_{\beta \gamma} +\left(\delta^\alpha_\beta \,\delta^\mu_{\gamma} +\delta^\alpha_\gamma \,\delta^\mu_\beta -g_{\beta \gamma} g^{\alpha \mu} \right)\,\lambda,_\mu \,,
\end{align}
and 
\begin{align} \label{Riem-tr2}
R^\mu{}_{\nu \alpha \beta} \to \widetilde{R}^\mu{}_{\nu \alpha \beta} =&  R^\mu{}_{\nu \alpha \beta}\, + \delta^\mu_\beta \lambda_{\nu \alpha} - \delta^\mu_\alpha \lambda_{\nu \beta} +g^{\mu \sigma}(g_{\nu\alpha}\lambda_{\sigma \beta}- g_{\nu\beta}\lambda_{\sigma \alpha}) \nonumber\\ & + (\delta^\mu_\beta g_{\nu\alpha} - \delta^\mu_\alpha g_{\nu\beta})%{\Box\lambda} 
g^{\sigma\tau}\lambda_{,\sigma}\lambda_{,\tau} \,,
\end{align}
respectively. In \eqref{Riem-tr2}, $\lambda_{\mu\nu}\equiv \nabla_\nu \lambda_{,\mu}
%\lambda,_{\mu};_{\nu} 
-\lambda,_{\mu}\lambda,_{\nu}$, where $\nabla$ is the covariant derivative with the Levi-Civita connection given by the Chistoffel symbol $\Gamma^\alpha_{\beta \gamma}$  (see \cite{Eisenhart} Chapter 28, p. 89). The  Ricci tensor ${R}_
{\nu\beta}=R^\mu{}_{\nu \mu \beta}$, and the Ricci scalar ${R}= R^\mu{}_{\nu\mu\beta}\,g^{\nu\beta}$ transform into:
\begin{align}
    {R}_
{\nu\beta} \to \tilde{R}_
{\nu\beta} ={R}_
{\nu\beta}+ (2-n) \lambda_
{\nu\beta}
+ (2-n)g_{\nu\beta}\,g^{\sigma\tau}\lambda_{,\sigma}\lambda_{,\tau} -g_{\nu\beta}\Box\lambda  \end{align}
and
\begin{align} \label{scaling-R}
\widetilde{R}= e^{-2\lambda}\left[{R} +(n-1)(2 \Box \lambda + (n-2) g^{\mu\nu} \lambda,_\mu \lambda,_\nu) \right]\,,     
\end{align}
respectively, where  $\Box \equiv g^{\mu \nu}\nabla_\mu\nabla_\nu$ is the D'Alembertian operator. 

If a scalar field transforms as in \eqref{s=0}, then its D'Alambertian transforms into
\begin{align}
\widetilde{\Box \phi} = e^{(p-2)\lambda}\Big[\Box \phi+ (p \Box \lambda + p[n+p-2]g^{\mu \nu} \lambda,_\mu \lambda,_\nu )\phi + (n+2p-2)g^{\mu \nu}\lambda,_\mu \phi,_\nu   \Big]\,.
\end{align}
Hence, for $p=(2-n)/2$ the term with the first derivative of $\phi$ drops out.

\section*{Acknowledgements}
 We thank interesting discussions and comments from P.~D. \'Alvarez, R. D'Auria, M.~A. Lled\'o, S. Maurelli, O. Mi\v{s}kovi\'c, R. Noris, L. Ravera, M. Trigiante. J.Z. acknowledges support from INFN and hospitality at the DISAT department of Politecnico di Torino. This work was also partially supported by ANID/FONDECYT grants 1230112, 1241835, 1252053 and 1262197.

%%%%%%%%%%%%%%%%%%%%%%%%%%%%%
\end{document}